\documentclass[twocolumn,secnumarabic,amssymb,nobibnotes,aps,pra,showpacs,superscriptaddress]{revtex4-1}
\usepackage[english]{babel}
\usepackage{graphicx}
\usepackage{amsmath}
\usepackage{amsfonts}
\usepackage[usenames,dvipsnames]{color}
\usepackage{hyperref}
\usepackage{blindtext}
\hypersetup{colorlinks}
\hypersetup{linkcolor=black, citecolor=black, urlcolor=blue}
\usepackage[utf8]{inputenc}
\usepackage{xcolor}
 \newcommand\mycolor{\color{black}\xspace}

\begin{document}
\title{Colloid--oil-water-interface interactions in the presence of multiple salts: charge regulation and dynamics}
\author{J. C. Everts}
\affiliation{Institute for Theoretical Physics, Center for Extreme Matter and Emergent Phenomena,  Utrecht University, Princetonplein 5, 3584 CC Utrecht, The Netherlands}
\email{jeffrey.everts@gmail.com}
\author{S. Samin}
\affiliation{Institute for Theoretical Physics, Center for Extreme Matter and Emergent Phenomena,  Utrecht University, Princetonplein 5, 3584 CC Utrecht, The Netherlands}
\author{N. A. Elbers}
\affiliation{Soft Condensed Matter, Debye Institute for Nanomaterials Science, Princetonplein 5, 3584 CC, Utrecht, The Netherlands}
\author{J. E. S. van der Hoeven}
\affiliation{Soft Condensed Matter, Debye Institute for Nanomaterials Science, Princetonplein 5, 3584 CC, Utrecht, The Netherlands}
\author{A. van Blaaderen}
\affiliation{Soft Condensed Matter, Debye Institute for Nanomaterials Science, Princetonplein 5, 3584 CC, Utrecht, The Netherlands}
\author{R. van Roij}
\affiliation{Institute for Theoretical Physics, Center for Extreme Matter and Emergent Phenomena,  Utrecht University, Princetonplein 5, 3584 CC Utrecht, The Netherlands}
\date{\today}

\begin{abstract}
We theoretically and experimentally investigate colloid-oil-water-interface interactions of
 charged, sterically stabilized, poly(methyl-methacrylate) colloidal particles dispersed in a low-polar oil (dielectric constant $\epsilon=5-10$) that is in contact with an adjacent water phase. In this model system, the colloidal particles cannot penetrate the oil-water interface due to repulsive van der Waals forces with the interface whereas the multiple salts that are dissolved in the oil are free to partition into the water phase. The sign and magnitude of the Donnan potential and/or the particle charge is affected by these salt concentrations such that the effective interaction potential can be highly tuned. Both the equilibrium effective colloid-interface interactions and the ion dynamics are explored within a Poisson-Nernst-Planck theory, and compared to experimental observations.
\end{abstract}

\maketitle

\section{Introduction}
Electrolyte solutions in living organisms often contain multiple ionic species 
such as $\text{Na}^+$, $\text{K}^+$, $\text{Mg}^{2+}$ and $\text{Cl}^{-}$. The 
concentrations of these ions and their affinity to bind to specific proteins 
determine the intake of ions from the extracellular space to the intracellular 
one \cite{Alberts}. In this example, the concentration of \emph{multiple} ionic 
species is used to \emph{tune} biological processes. However, this scenario is 
not limited to living systems, as it can also be important for ionic liquids 
\cite{Chatel:2014}, batteries \cite{Guerfi:2010}, electrolytic cells 
\cite{Westbroek:2015}, and colloidal systems \cite{Elbers:2016, Ninathesis, 
Jessi}, as we  show in this paper. 

In colloidal suspensions, the dissolved salt ions screen the surface charge of 
the colloid, leading to a monotonically decaying diffuse 
charge layer in the fluid phase. At the same time, these ions may adsorb to 
the colloid surface and modify its charge \cite{lyklema1995}. The colloid 
surface may also possess multiple ionizable surface groups that respond to the 
local physico-chemical conditions \cite{Ninham:1971,prieve1976}. Hence, the 
particle charge is determined by the ionic strength of the medium and the
particle distance from other charged interfaces. This so-called charge regulation is 
known to be crucial to correctly describe the interaction between charged 
particles in aqueous solutions, from nanometer-sized proteins 
\cite{warshel2006} to micron-sized colloids \cite{trefalt2016}. 

Colloidal particles are also readily absorbed at fluid-fluid interfaces,
such as air-water and oil-water interfaces, since this leads to a 
large reduction in the surface free energy, of the order of $10^5-10^7k_BT$ 
per particle, where $k_BT$ is the thermal energy \cite{Pieranski:1980}. In an 
oil-water mixture, colloids therefore often form Pickering emulsions that consist of particle-laden droplets \cite{Pickering:1907, binks2006}, which have been the 
topic of extensive research due to their importance in many industrial processes, such as biofuel upgrade \cite{crossley2010}, 
crude oil refinery \cite{binks2006}, gas storage 
\cite{carter2010}, and as anti-foam agents \cite{denkov2004}.

When the colloidal particles penetrate the fluid-fluid interface, the 
electrostatic component of the particle-particle interactions is modified by the dielectric 
mismatch between the fluid phases \cite{hurd1985}, nonlinear 
charge renormalization effects \cite{Frydel:2007,Frydel:2011}, and the different 
charge regulation mechanisms in each phase \cite{Choi:2014}. The resulting long 
range lateral interactions have been studied in detail \cite{Pieranski:1980,hurd1985,Frydel:2007,Frydel:2011}, with the out-of-plane 
interactions also receiving some attention \cite{Zwanikken:2007, 
Leunissen2:2007}. Less attention has been dedicated to the electrostatics of 
the particle-interface interaction, although it is essential for understanding 
the formation and stability of Pickering emulsions.

In this work, we focus on an oil-water system, where oil-dispersed charged and 
sterically stabilized poly(methyl-methacrylate) (PMMA) particles are found 
to be trapped near an oil-water interface, \emph{without} penetrating it, due 
to a force balance between a repulsive van der Waals (vdW) and an attractive 
image-charge force between the colloidal particle and the interface \cite{parsegian, 
Oettel:2007, Elbers:2016}. Here, the repulsive vdW forces stem from the particle dielectric constant that is smaller than that of water and oil. {\mycolor{This can be understood from the fact that for the
three-phase system of PMMA-oil-water, the difference in dielectric spectra determine whether the vdW interaction is attractive or repulsive \cite{parsegian}, while for two-phase systems, like atoms in air, the vdW interaction is always attractive.}}
In addition to this force balance, we have recently 
shown in Ref. \cite{Everts:2016b} that the dissolved ions play an important 
role in the emulsion stability. In addition to the usual screening and charge regulation, ions can 
redistribute among the oil and water phase according to their solvability and 
hence generate a charged oil-water interface that consists of a back-to-back electric double layer. Within a 
single-particle picture, this ion partitioning can be shown to modify the 
interaction between the colloidal particle and the oil-water interface. For a 
non-touching colloidal particle, the interaction is tunable from attractive to 
repulsive for large enough separations, by changing the sign of the product 
$Z\phi_D$ \cite{Everts:2016b}, where $Ze$ is the particle charge and $k_BT\phi_D/ e$ the Donnan 
potential between oil and water due to ion partitioning, with $e$ the 
elementary charge. The tunability of colloid-ion forces is a central theme of 
this work, in which we will explore how the quantities $Z$ and $\phi_D$ can be 
rationally tuned. 

Although tuning the interaction potential through $Z\phi_D$ is quite general, the salt concentrations in a binary mixture of particle-charge determining positive and negative ions cannot be varied independently due to bulk charge neutrality; in other words, $Z\phi_D$ is always of a definite sign for a given choice of two ionic species. This motivates us to extend the formalism of Ref. \cite{Everts:2016b} by including at least three ionic species which are all known to be present in the experimental system of interest that we will discuss in this paper. Including a second salt compound with an ionic species common to the two salts, allows us to independently vary the ionic strength and the particle charge. Because of this property, it is then possible to tune the sign of the particle charge, which is acquired by the ad- or de-sorption of ions, via the salt concentration of one of the two species. Furthermore, for more than two types of ions, the Donnan potential depends not only on the difference in the degree of hydrophilicity between the various species \cite{Bier:2008}, but also on the bulk ion concentrations \cite{Westbroek:2015}. This leads to tunability of the magnitude, and possibly the sign, of the Donnan potential.

We apply our theory to experiments, where seemingly trapped colloidal particles near an oil-water interface could surprisingly be detached by the addition of an organic salt to the oil phase \cite{Elbers:2016}. We will show that our minimal model including at least three ionic species is sufficient to explain the experiments. We do this by investigating the equilibrium properties of the particle-oil-water-interface effective potential in presence of multiple salts and by examining out-of-equilibrium properties, such as diffusiophoresis.  The latter is relevant for recent experiments where diffusiophoresis was found to play a central role in the formation of a colloid-free zone at an oil-water interface \cite{Florea:2014,Musa:2016,Squiresa:2016}.

As a first step, we set up in Sec. \ref{sec:DF} the density functional for the model system. In Sec. \ref{sec:systemexp}, the experiments are described. In Sec. \ref{sec:collint}, the equilibrium effective colloid-interface interaction potentials are explored as function of salt concentration, and we work out a minimal model that can account for the experimental observations. In  Sec. \ref{sec:PNP}, we look at the influence of the ion dynamics within a Poisson-Nernst-Planck approach, and investigate how the system equilibrates when no colloidal particle is present. We conclude this paper by elucidating how our theory compares against the experiments of Elbers \emph{et al.} \cite{Elbers:2016}, where multiple ionic species were needed to detach colloidal particles from an oil-water interface.

\section{Density functional}
\label{sec:DF}
Consider two half-spaces of water ($z<0$, dielectric constant $\epsilon_w=80$) and oil ($z>0$, dielectric constant $\epsilon_o$) at room temperature $T$ separated by an interface at $z=0$. We approximate the dielectric constant profile by $\epsilon(z)=(\epsilon_o-\epsilon_w)\Theta(z)+\epsilon_w$, with $\Theta(z)=[1+\tanh(z/2\xi)]/2$ and $\xi$ the interface thickness. Since we take $\xi$ to be molecularly small, we can interpret $\Theta$ as the Heaviside step function within the numerical accuracy on the micron length scales of interest here. The $N_+$ species of monovalent cations and $N_-$ species of monovalent anions can be present as free ions in the two solvents, and are described by density profiles $\rho_{i,\alpha}({\bf r})$ ($i=1,...,N_\alpha$, $\alpha=\pm$) with bulk densities in water (oil) $\rho_{i,\alpha}^w$ ($\rho_{i,\alpha}^o$). Alternatively, the ions can bind to the surface of a charged colloidal sphere
(dielectric constant $\epsilon_c$, radius $a$, distance $d$ from the interface) with areal density $\sigma_{i,\alpha}({\bf r})$. The colloidal surface charge density $e\sigma({\bf r})$ is thus given by $\sigma({\bf r})=\sum_{i=1}^{N_+} \sigma_{i,+}({\bf r})- \sum_{i=1}^{N_-}\sigma_{i,-}({\bf r})$. The ions can partition among water and oil, which is modeled by the external potentials $V_{i,\alpha}(z)=\beta^{-1}f_{i,\alpha}\Theta(z)$ (where $\beta^{-1}=k_BT$), where the self-energy $f_{i,\alpha}$ is defined as the (free) energy cost to transfer a single ion from the water phase to the oil phase. 

The effects of ion partitioning and charge regulation can elegantly be captured within the grand potential functional, $\Omega\left[\{\rho_{i,\pm},\sigma_{i,\pm}\}_{i=1}^{N_\pm}; d\right]$, given by
\begin{align}
\Omega=&\ \mathcal{F} -\sum_{\alpha=\pm}\sum_{i=1}^{N_\alpha}\int d^3{\bf r}\Bigl\{\big[\mu_{i,\alpha}\!-\!V_{i,\alpha}(z)\big]\nonumber \\&\times\big[\rho_{i,\alpha}({\bf r})+\sigma_{i,\alpha}({\bf r})\delta(|{\bf r}\!-\!d{\bf e}_z|\!-\!a)\big]\Bigr\}, 
\end{align}
with $\mu_{i,\alpha}=k_B T\ln({\rho_{i,\alpha}^w}\Lambda_{i,\alpha}^3)$ the chemical potential of the ions in terms of the ion bulk concentrations $\rho_{i,\alpha}^w$ in water and ${\bf e}_z$ the normal unit vector of the planar interface.  Here the Helmholtz free energy functional $\mathcal{F}$ is given by
\begin{widetext}
\begin{align}
\beta\mathcal{F}&\left[\{\rho_{i,\pm},\sigma_{i,\pm}\}_{i=1}^{N_\pm}; d\right]=\sum_{\alpha=\pm}\sum_{i=1}^{N_{\alpha}}\int_\mathcal{R}d^3{\bf r} \ \rho_{i,\alpha}({\bf r})\left\{\ln\left[\rho_{i,\alpha}({\bf r})\Lambda_{i,\alpha}^3\right]-1\right\}+\frac{1}{2}\int_\mathcal{R}d^3{\bf r}\ Q({\bf r})\phi({\bf r}) \nonumber \\
&+\sum_{\alpha=\pm}\sum_{i=1}^{N_\alpha}\int_\Gamma d^2{\bf r} \ \Bigg({}\sigma_{i,\alpha}({\bf r})\left\{\ln [\sigma_{i,\alpha}({\bf r})a^2]+\ln\left( K_{i,\alpha}\Lambda_{i,\alpha}^3\right)\right\}+\left[\sigma_m\theta_{i,\alpha}-\sigma_{i,\alpha}({\bf r})\right]\ln\left\{\left[\sigma_m\theta_{i,\alpha}-\sigma_{i,\alpha}({\bf r})\right]a^2\right\}\Bigg), 
\label{eq:dft}
\end{align}
\end{widetext}
where the region outside the colloidal particle is denoted by $\mathcal{R}$ and the particle surface is denoted by $\Gamma$. The first term of Eq. \eqref{eq:dft} is an ideal gas contribution. The mean-field electrostatic energy is described by the second term of Eq. \eqref{eq:dft} which couples the total charge density $Q({\bf r})=\sum_{i=1}^{N_+} \rho_{i,+}({\bf r})-\sum_{i=1}^{N_-} \rho_{i,-}({\bf r})+\sigma({\bf r})\delta(|{\bf r}-d{\bf e}_z|-a)$ to the electrostatic potential $\phi({\bf r})/\beta e=25.6 \ \phi({\bf r}) \ \text{mV}$. The final term is the free energy of an $(N_++N_-+1)$-component lattice gas of neutral groups and charged groups, with a surface density of ionizable groups $\sigma_m a^2=10^6$  (or one ionizable group per $\text{nm}^2$) and $\theta_{i,\alpha}$ is the fraction of ionizable groups available for an ion of type $(i,\alpha)$. A neutral surface site S$_{i,\alpha}$ can become charged via adsorption of an ion $\text{X}_i^{i, \alpha}$, i.e., $\text{S}_{i,\alpha}+\text{X}_{i,\alpha}^{\alpha}\leftrightarrows \text{S}_{i,\alpha}\text{X}_{i,\alpha}^{\alpha}$ with an equilibrium constant $K_{i,\alpha}=[\text{S}_{i,\alpha}][\text{X}_{i,\alpha}^{\alpha}]/[\text{S}_{i,\alpha}\text{X}_{i,\alpha}^{\alpha}]$ and $pK_{i,\alpha}=-\log_{10}(K_{i,\alpha}/1\ \text{M})$. 

From the Euler-Lagrange equations $\delta\Omega/\delta\rho_{i,\alpha}({\bf r})=0$, we find the equilibrium profiles $\rho_{i,\pm}({\bf r})=\rho_{i,\pm}^w\exp[\mp\phi({\bf r})+f_{i,\alpha}\Theta(z)]$.
Combining this with the Poisson equation for the electrostatic potential, we obtain the Poisson-Boltzmann equation for ${\bf r}\in\mathcal{R}$,
\begin{equation}
\nabla\cdot[\epsilon(z)\nabla\phi({\bf r})]/\epsilon_o=\kappa(z)^2\sinh[\phi({\bf r})-\Theta(z)\phi_D],
\label{eq:PB}
\end{equation}
where we used bulk charge neutrality to find the Donnan potential $\phi_D/\beta e$ given by,
\begin{equation}
\phi_D=\frac{1}{2}\log\left[\frac{\sum_i\rho_{i,+}^w\exp(-f_{i,+})}{\sum_i\rho_{i,-}^w\exp(-f_{i,-})}\right]. \label{eq:donnan}
\end{equation}
In Eq. \eqref{eq:PB}, we also introduced the inverse length scale $\kappa(z)=\sqrt{8\pi\lambda_B^o\rho_s(z)}$, with
\begin{equation} 
\rho_s(z)=\frac{1}{2}\sum_{\alpha=\pm}\sum_{i=1}^{N_\alpha}\rho_{i,\alpha}^o\exp[(\alpha\phi_D+f_{i,\alpha})\Theta(-z)],
\end{equation} where the Bjerrum length in oil is given by $\lambda_B^o=e^2/4\pi\epsilon_\text{vac}\epsilon_o k_BT$. Notice that $\kappa(z)=\kappa_o$ for $z>0$, with $\kappa_o^{-1}$ the screening length in oil, and that for $z<0$ we have that $\kappa(z)=\kappa_w\sqrt{\epsilon_w/\epsilon_o}$, with $\kappa_w^{-1}$ the screening length in water. Finally, the bulk oil densities $\rho_{i,\alpha}^o$ are related to the bulk water densities as
\begin{equation}
\rho_{i,\alpha}^w=\rho_{i,\alpha}^o\exp[(\alpha\phi_D+f_{i,\alpha})]. \label{eq:salt}
\end{equation}
Inside the dielectric colloidal particle, the Poisson equation reads $\nabla^2\phi=0$. On the particle surface, ${\bf r}\in\Gamma$, we have the boundary condition 
${\bf n}\cdot[\epsilon_c\nabla\phi|_\text{in}-\epsilon_o\nabla\phi|_\text{out}]/\epsilon_o=4\pi\lambda_B^o\sigma({\bf r})$,
with a charge density described by the Langmuir adsorption isotherm for ${\bf r}\in\Gamma$,
\begin{equation}
\sigma_{i,\alpha}({\bf r})=\frac{\sigma_m\theta_{i,\alpha}}{1+(K_{i,\alpha}/\rho_{i,\alpha}^o)\exp\{\alpha[\phi({\bf r})-\phi_D]\}},\label{eq:Langmuir}
\end{equation}
which follows from $\delta\Omega/\delta\sigma_{i,\alpha}({\bf r})=0$. 

Eqs.  \eqref{eq:PB}-\eqref{eq:Langmuir} are solved numerically for $\phi({\bf r})$ using the cylindrical symmetry, and generic solutions were already discussed in the case of a single adsorption model in Ref. \cite{Everts:2016b}. From the solution we determine $\rho_{i,\alpha}({\bf r})$ and $\sigma_{i,\alpha}({\bf r})$. These in turn determine the effective colloid-interface interaction Hamiltonian via
\begin{equation}
H(d)=\Phi_\text{VdW}(d)+\underset{\{\rho_{i,\pm},\sigma_{i,\pm}\}_{i=1,...,N_\pm}}\min\Omega\left[\{\rho_{i,\pm},\sigma_{i,\pm}\}_{i=1}^{N_\pm}; d\right]. \label{eq:blabla}
\end{equation} Here, we added the vdW sphere-plane potential $\Phi_\text{VdW}$, with an effective particle-oil-water Hamaker constant $A_H$ \cite{parsegian}. Eq. \eqref{eq:blabla} can then be evaluated to give 
\begin{widetext}
\begin{align}
\beta H(d)&=\int_\mathcal{R}\ d^3{\bf r} \ \rho_s(z)\Big\{\phi({\bf r})\sinh[\phi({\bf r})-\Theta(z)\phi_D]-2(\cosh[\phi({\bf r})-\Theta(z)\phi_D]-1)\Big\}-\frac{1}{2}\int_\Gamma d^2{\bf r} \ \sigma({\bf r})\phi({\bf r})\label{eq:effHam} \\
&-\sum_{\alpha=\pm}\sum_{i=1}^{N_\alpha}\sigma_m\theta_{i,\alpha}\int_\Gamma d^2{\bf r}\ln\left(1+\frac{\rho_{i,\alpha}^o}{K_{i,\alpha}}\exp\{-\alpha[\phi({\bf r})-\phi_D]\}\right)-\frac{\beta A_H}{6}\left[\frac{1}{d/a-1}+\frac{1}{d/a+1}+\ln\left(\frac{d/a-1}{d/a+1}\right)\right], \nonumber
\end{align}
\end{widetext}
which we will investigate using the experimental parameters given in Table \ref{tab:system}, to be elucidated in the next section.
\\
\begin{table}[h]
\caption{System parameters (symbols explained in main text)}
\centering
\begin{tabular}{l*{5}{c}}
\hline
System              & $\epsilon_o$ & $\eta$ & $Z$ & $\rho_{\text{TBA}^+}\big{|}_{Z=0}$[$\mu$M] & $(\kappa_o\big{|}_{Z=0})^{-1}$[$\mu$m]  \smallskip\\
\hline
1 & 7.92 & 0.01 & +930  & 1-5 & $\sim 1$ \\
2 & 6.2 & 0.01 & -280  & n.a. & n.a.  \\
\hline
\end{tabular}
\label{tab:system}
\end{table}
\begin{figure*}
\includegraphics[width=0.7\textwidth]{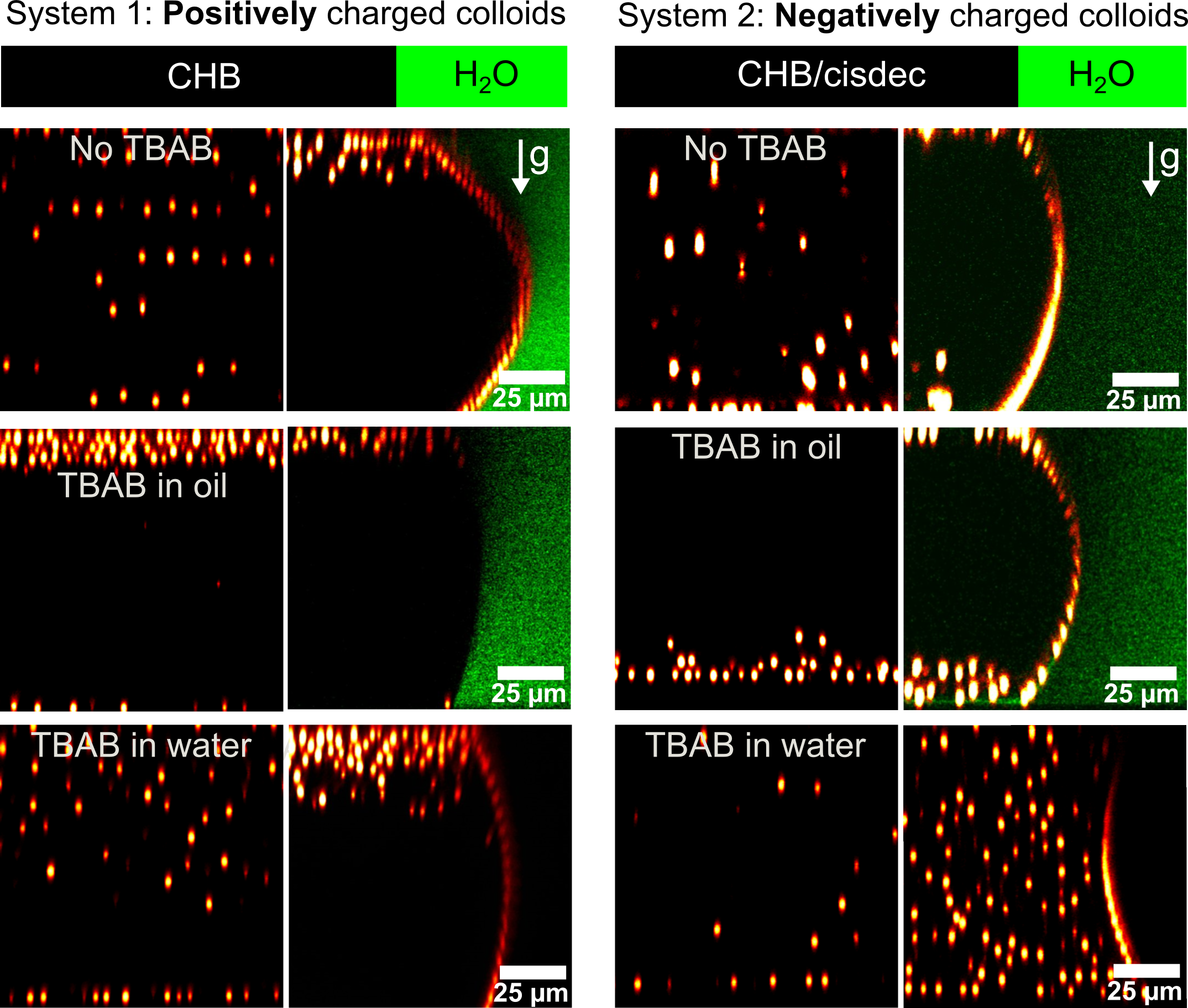}
\centering
\caption{Confocal images of positively (system 1, $Z= +930$) and negatively charged (system 2, $Z= -280$) PMMA colloidal particles close to an oil-water interface in the absence (top) and presence of salt in the oil (middle) and water phase (bottom). In system 1 the particles detach from the interface upon addition of 300 $\mu$M TBAB to the oil phase, whereas no detachment was observed upon addition of 300 $\mu$M TBAB to the oil phase in system 2. In both systems the colloidal particles were attracted to the interface when adding 50 mM TBAB to the water phase. The oil phase in system 1 and 2 consists of pure CHB and a mixture of CHB/27.2~wt\% cis-decalin, respectively. {\mycolor For the TBAB in water experiments we did not use the dye FITC (green) to colour the water phase to prevent the formation of excess charges due to interactions between FITC and TBAB.}}
\label{fig:exp}
\end{figure*}
\begin{figure}[ht]
\includegraphics[width=0.5\textwidth]{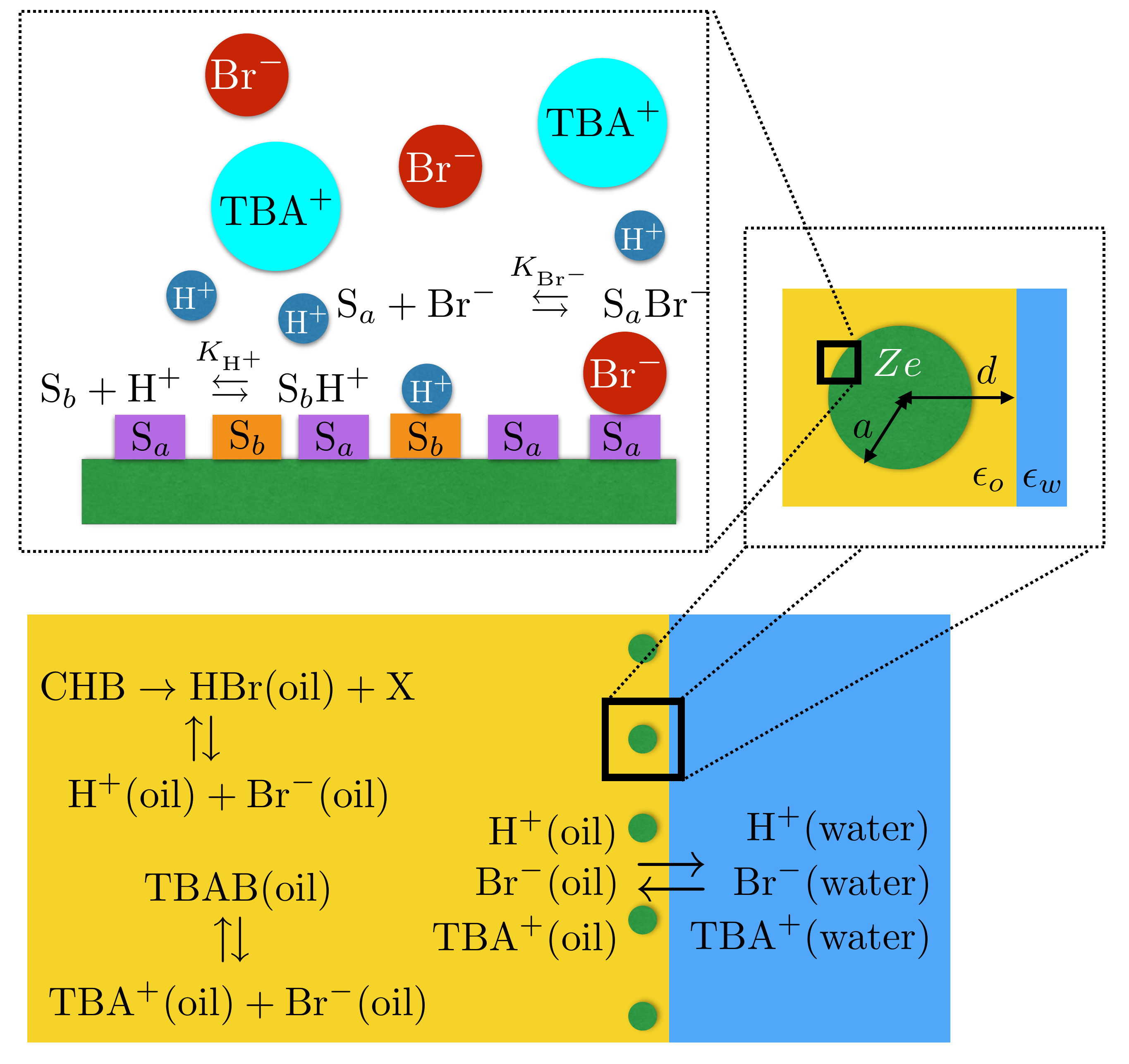}
\caption{Schematic overview of the experiment of hydrophobic colloidal particles in a demixed CHB-water system. The oily solvent CHB decomposes giving rise to HBr and a decomposition product X, with HBr in equilibrium with $\mathrm{H}^+$ and $\mathrm{Br}^-$ ions. In addition to HBr, we also include the organic salt TBAB, so that the concentration of $\mathrm{Br}^-$ is not necessarily equal to the $\mathrm{H}^{+}$ concentration. TBAB is in equilibrium with free $\mathrm{TBA}^+$ and $\mathrm{Br}^-$ ions, and all ions can partition between water and oil. For simplicity we do not take into account the equilibria between the undissociated salts and the free ions. We consider a colloidal particle with radius $a$ and at a distance $d$ from the interface has a charge $Ze$. On the particle surface there are two types of binding sites, $\mathrm{S}_a$ and $\mathrm{S}_b$, of which the former can bind a $\mathrm{Br}^-$ ion with equilibrium constant $K_{\text{Br}^-}$ and the latter can bind $\mathrm{H}^+$ ion with equilibrium constant $K_{\text{H}^+}$. We assume that $\mathrm{TBA}^+$ cannot bind to the particle surface. There are also colloidal particles in the bulk that are separated from the two-dimensional monolayer at the interface by a zone void of colloids (not shown for clarity).}
\label{fig:schematic}
\end{figure}

\section{System and experimental observations}
\label{sec:systemexp}

We consider two experimental systems from Ref. \cite{Elbers:2016}, to which we 
will refer as system 1 and 2. Both systems are suspensions with sterically 
stabilized poly(methyl-methacrylate) (PMMA) colloidal particles of radius $a=1.4\ 
\mu\mathrm{m}$ and dielectric constant $\epsilon_c=2.6$ \cite{Leunissen:2007, 
Elbers:2016}. The comb-graft steric stabilizer is composed of 
poly(12-hydroxystearic acid) (PHSA) grafted on a backbone of PMMA 
\cite{Bosma:2002, Elsesser:2010}. This stabilizer was covalently bonded to the 
particles in system 1 (resulting in so-called locked PMMA particles 
\cite{Linden:2015}) whereas it was adsorbed to the surface of the particles in 
system 2 (resulting in so-called unlocked PMMA particles \cite{Linden:2015}). 
{\mycolor{In the locking process the PMMA colloids acquire a higher surface potential and charge. The increase in charge is mainly due to the incorporation of 2-(dimethylamino)ethanol in the PMMA colloids during the locking procedure. The protonation of the incorporated amine groups renders colloidal particles with an increased positive charge (see also Ref. \cite{Linden:2015}). Locked particles (like in system 1) are thus always positively charged and can only become negative by introducing TBAB. Unlocked particles can be either (slightly) positively or negatively charged.}}

The locked particles in system 1 were dispersed in deionized cyclohexylbromide 
(CHB) and were positively charged, whereas the unlocked particles in system 2 
were dispersed in CHB/cis-decalin (27.2 wt\%) and were negatively charged. The 
key parameters of both systems are summarized in Table \ref{tab:system}, where 
$\eta$ is the volume fraction. It is important to note that CHB decomposes  in 
time, producing HBr. Since CHB is a non-polar oil ($\epsilon_o=5-10$), rather than an apolar oil ($\epsilon_o\approx2$), which means that the 
dielectric constant is high enough for significant dissociation of (added) 
salts to occur, specifically, HBr can dissociate into $\text{H}^+$ and 
$\text{Br}^-$ ions, which can subsequently adsorb on the particle surface 
\cite{Linden:2015}. In an oil phase without added salt {\mycolor{and without an adjacent water phase}}, $\kappa_o^{-1}=6 \ 
\mu\mathrm{m}$ was assumed for both systems \cite{Linden:2015}, {\mycolor{which is a reasonable estimate based on conductivity measurements or the crystallization behaviour of colloidal particles dispersed in CHB}}.

In the experimental study suspensions of system 1 and 2 were brought in borosilicate capillaries (5 cm $\times$ 2.0 mm $\times$ 0.10 mm) which were already half-filled with deionized water; the colloidal behavior near the oil-water interface was studied with confocal microscopy. {\mycolor When necessary the oil-water interface was more clearly visualized by using FITC-dyed water instead of ultrapure water. FITC water was taken from a stock solution to which an excess of FITC dye was added. FITC water was never used in combination with TBAB in the aqueous phase to prevent interactions between the FITC dye and TBAB.} In Fig. \ref{fig:exp}, the confocal images of both systems before (top) and after addition of organic salt tetrabutylammoniumbromide (TBAB) to the oil (middle) and water phase (bottom) are shown. In the absence of salt, the force balance between image charge attractions and vdW repulsion leads to the adsorption of the colloidal particles at the interface in both systems \cite{Oettel:2007}, without the colloidal particles penetrating the oil-water interface \cite{Elbers:2016}. In addition, the water side of the interface was reported to be positively charged, while the oil side is negatively charged ~\cite{Leunissen:2007}. When TBAB was added to the oil phase above the threshold concentration $\rho_{\text{TBA}^+}\big{|}_{Z=0}$ mentioned in table \ref{tab:system}, with corresponding Debye screening length in oil $(\kappa_o\big{|}_{Z=0})^{-1}$, the colloidal particles in system 1 were driven from the interface towards the bulk oil phase, whereas the addition of TBAB did not result in particle detachment in system 2, see Fig. \ref{fig:exp}. Over time the detached colloidal particles in system 1 reattached close to the oil-water interface \cite{Elbers:2016} (see Fig. S1 in the supplemental information). When TBAB was added to the water phase, the colloidal particles in both system 1 and 2 were driven from the bulk oil to the oil-water interface, producing dense layers of colloidal particles near the interface \cite{Elbers:2016}, see Fig. \ref{fig:exp} and Fig. S2 in the supplemental information. Finally, we also investigated system 1 under the same density-matching conditions as in system 2, and observed no qualitative change in the response to salt addition, see Fig. S3 in the supplemental information. 

When the TBAB was added to the oil phase, the positively charged colloidal particles in system 1 reversed the sign of their charge $Z=\int_\Gamma d^2{\bf r}\ \sigma({\bf r})$ from positive ($Z>0$) to negative ($Z<0$) \cite{Elbers:2016}. This suggests that $\text{H}^+$ and $\text{Br}^-$ can both adsorb to the particle surface and that the addition of TBAB introduces more $\text{Br}^-$ in the system, causing the particle charge of system 1 to become negative for a high enough concentration of TBAB. The estimated concentration of free TBA$^+$ ions $\rho_{\text{TBA}^+}\big{|}_{Z=0}$, and the corresponding Debye length $(\kappa_o\big{|}_{Z=0})^{-1}$ in our experiments are listed in Table \ref{tab:system}. Both parameters are not defined for system 2 (not applicable, n.a.), since here negative particles cannot become positively charged in the setup that we consider, {\mycolor{because we always observed that adding TBAB results in a more negative particle charge}}. In Fig. \ref{fig:schematic}, all equilibria, including the decomposition of CHB, the equilibria of HBr and TBAB with their free ions, and the partitioning of these ions between water and oil, are schematically shown. For simplicity, we have not taken the salt decomposition equilibria into account in the theory of Sec. \ref{sec:DF}. However, the Bjerrum pairs HBr and TBAB could be included in the theory by using the formalism of Ref. \cite{Valeriani:2010}. In the upper right inset of Fig. \ref{fig:schematic}, we show schematically the binding of $\text{H}^+$ and $\text{Br}^-$ onto the particle surface. In principle, TBA$^+$ can also adsorb on the particle surface, but we expect this to be a small effect that we neglect. {\mycolor{This is justified since adding TBAB renders more negative particles, suggesting that $\text{Br}^-$ can more easily adsorb on the particle surface than TBA$^+$. Hence, including a finite value for $K_{\text{TBA}^+}$ in our model does not change our results qualitatively, but only quantitatively.}}

We will explain the experimental observations described in this section by applying the formalism of section \ref{sec:DF}. Moreover, we will discuss the differences between a single adsorption model and a binary adsorption model and the influence of a third ionic species, which is a first extension trying to get closer to the full experimental complexity compared to our previous work \cite{Everts:2016b}, where only a single adsorption model was considered in a medium with only two ionic species.

\section{Colloid-interface interactions}
\label{sec:collint}
We will perform calculations for up to two species of cations ($N_+=1,2$) and one species of anions $N_-=1$, where $(1,+)$ corresponds to $\text{H}^+$, $(1,-)$ to $\text{Br}^-$, and $(2,+)$ to $\text{TBA}^+$. To estimate the order of magnitude of the ion sizes, we consider their effective (hydrated) ionic radii $a_{\text{H}^+}=0.28 \ \text{nm}$, $a_{\text{Br}^-}=0.33\ \text{nm}$, and $a_{\text{TBA}^+}=0.54 \ \text{nm}$ \cite{Leunissen2:2007}. This gives self-energies (in units of $k_BT$): $f_{\text{H}^+}=11$, $f_{\text{Br}^-}=10$ and $f_{\text{TBA}^+}=6$, based on the Born approximation $f_\alpha=(\lambda_B^o/2 a_\alpha)(1-\epsilon_o/\epsilon_w)$. This is a poor approximation in the case of $\text{TBA}^+$, because it is known that $\text{TBA}^+$ is actually a hydrophobic ion, $f_{\text{TBA}^+}<0$. However, this simple approximation does not affect our predictions since we can deduce from Eq. \eqref{eq:donnan} the inequality $(f_{\text{Br}^-}-f_{\text{TBA}^+})/2\leq\phi_D\leq(f_{\text{Br}^-}-f_{\text{H}^+})/2$. Therefore, as long as $f_{\text{TBA}^+}<f_{\text{Br}^-}$, we find that the Donnan potential is varied between a negative value and a positive one by adding TBAB, in line with experimental observations. Setting $f_{\text{TBA}^+}<0$ is therefore not required. Since we will fix $\kappa_o^{-1}$ throughtout our calculations, assuming $f_{\text{TBA}^+}<0$ would only affect the value of $\kappa_w^{-1}$, and we have already shown in our previous work that this parameter is not important for the colloid-interface interaction of oil-dispersed colloidal particles \cite{Everts:2016b}. We therefore use the Born approximation to analyze the qualitative behaviour of the effective interactions, such that $\phi_D$ can vary between $-0.5$ and $2$.

In an isolated oil phase {\mycolor{without an adjacent water phase}}, the screening length in our experiments was approximated to be $\kappa_o^{-1}=6\ \mu\mathrm{m}$. However, $\kappa_o^{-1}$ becomes larger in the presence of an adjacent water phase, since water acts as an ion sink: the ions dissolve better in water than in oil and therefore diffuse towards the water phase. The charged colloidal particles in the oil phase will counteract this effect, because these colloidal particles are always accompanied by a diffuse ion cloud, keeping some of the ions in the oil. Because we do not know the exact value of $\kappa_o^{-1}$ in an oil-water system, we consider it as a free parameter and let it vary in a reasonable range between $6\ \mu\mathrm{m}$ and $50\ \mu\mathrm{m}$. In our single-particle picture, we neglect many-body effects which can reduce the value of $\kappa_o^{-1}$, due to the overlap of double layers. This can be taken into account by introducing an effective Debye length \cite{Trizac:2002, Zoetekouw:2006, Zoetekouw2:2006}. Another many-body effect that we do not include, is the discharging of particles when the particle density is increased \cite{Vissers2:2011}. One should keep this in mind when directly comparing the values we use for $\kappa_o^{-1}$ to experiment.

\begin{figure*}[t]
\centering
\includegraphics[width=0.8\textwidth]{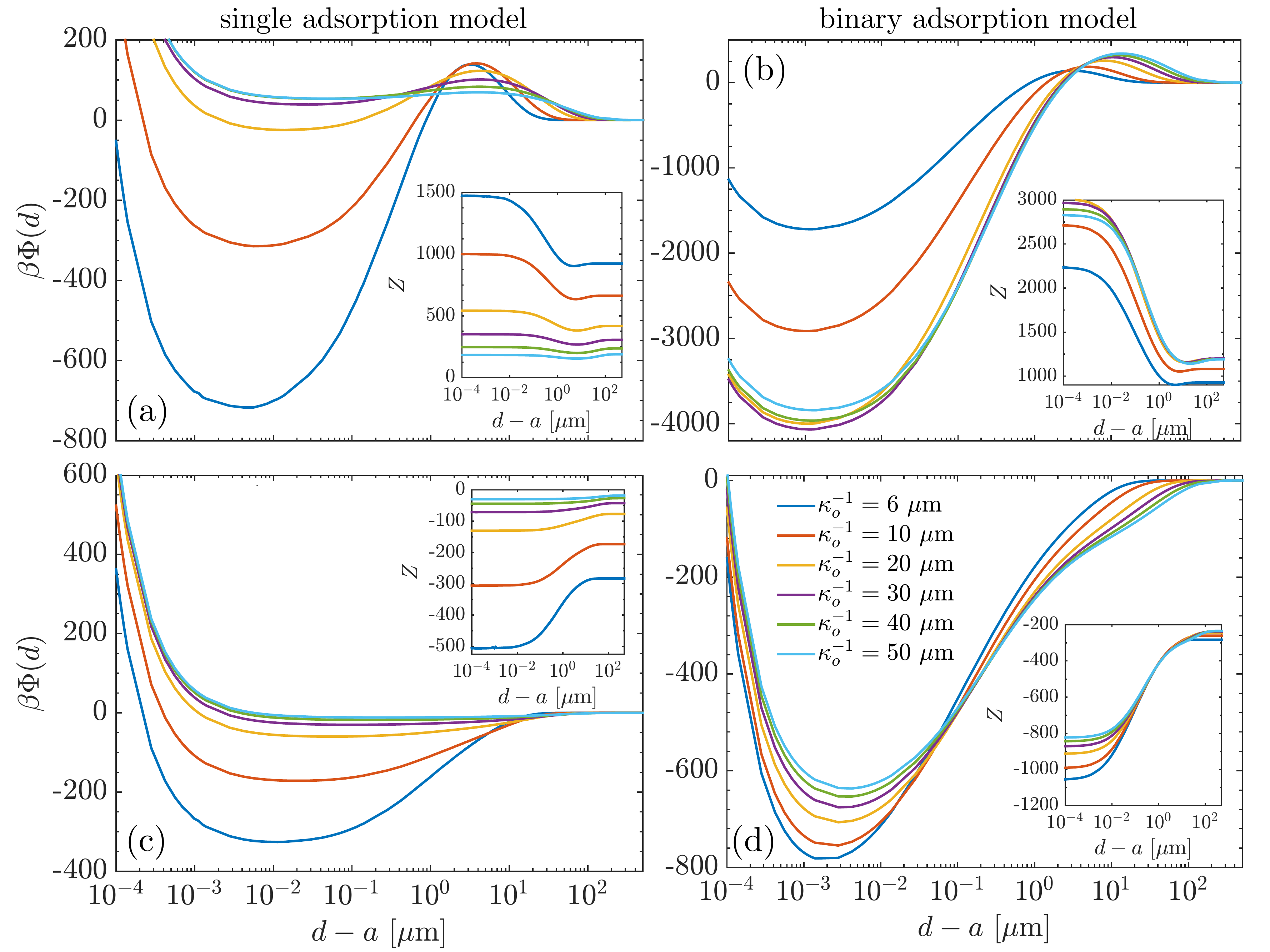}
\caption{The colloid-interface interaction potential {\mycolor{$\Phi(d)$ in units of $\beta^{-1}=k_BT$}} for the two systems in Table \ref{tab:system} for different oil Debye lengths $\kappa_o^{-1}$, as indicated by the legend in panel (d). Insets: The corresponding particle charges $Z(d)$. In System 1 the colloidal particles are positively charged, while in system 2 they are  negatively charged. We determine the equilibrium constants by matching the charges $Z$ at $d\rightarrow\infty$ to the values of $Z$ in Table \ref{tab:system} for Debye length in oil $\kappa_o^{-1}=6 \ \mu\mathrm{m}$. For system 1, we consider (a) a single adsorption model where only $\mathrm{H}^+$ can attach to the particle surface with equilibrium constant $a^3K_{\mathrm{H}^+}=165$, while $K_{\text{Br}^-}\rightarrow\infty$ ($\mathrm{Br}^-$ cannot adsorb) and (b) a binary adsorption model with $a^3K_{\mathrm{H}^+}=0.0001$, $a^3K_{\text{Br}^-}=47$ and fraction of sites available for $\text{Br}^-$, $\theta=0.8$. For the determination of these values we also used the salt concentration at which $Z$ switches sign by the addition of TBAB. We do similar calculations for system 2, for the single-adsorption model in (c), where we assume that no $\mathrm{H}^+$ can adsorb, but $\mathrm{Br}^-$ can with $a^3K_{\text{Br}^-}=3310$ and (d) in the binary adsorption model we take $a^3K_{\mathrm{H}^+}=1$ and $a^3K_{\text{Br}^-}=0.055$, with $\theta=0.5$. The Hamaker constant for the particle--oil-water interface vdW interaction in all panels is $\beta A_H=-0.3$.}
\label{fig:collint1}
\end{figure*}

\subsection{Systems without TBAB added}
\label{sec:eqnotbab}

In this subsection, we first investigate systems without the added TBAB (such that H$^+$ and Br$^-$ are the only ionic species) for two different adsorption models. The first one is a single-ion adsorption model. In this case, system 1 in Table I is described by the adsorption of $\text{H}^+$ alone, while for system 2 only $\text{Br}^-$ can adsorb. We use the experimental values of $Z$ from Table \ref{tab:system} to determine the equilibrium constants on the basis of a spherical-cell model in the dilute limit with $\kappa_o^{-1}=6\ \mu\mathrm{m}$. {\mycolor{Note that these values are obtained for colloidal particles dispersed in CHB \emph{without} an adjacent water phase.}} Within this procedure, we find $a^3K_{\mathrm{H}^+}=165$ and $K_{\text{Br}^-}\rightarrow\infty$ for system 1, while for system 2 we find $a^3K_{\text{Br}^-}=3310$ and $a^3K_{\mathrm{H}^+}\rightarrow\infty$. For the particle--oil-water-interface vdW interaction we use a Hamaker constant $\beta A_H=-0.3$, {\mycolor{which is an estimate based on the Lifshitz theory for the vdW interaction}} \cite{Elbers:2016}. The resulting colloid-interface interaction potentials as function of $\kappa_o^{-1}$ are shown in Fig. \ref{fig:collint1}(a) and (c), with the corresponding $Z(d)$ in the inset. The product $Z\phi_D$ determines the long-distance nature of the colloid-interface interaction: in Fig. \ref{fig:collint1}(a) it is repulsive for system 1, since $Z\phi_D<0$ and in Fig. \ref{fig:collint1}(c) attractive for system 2, since $Z\phi_D>0$ (recall that here $\phi_D=-0.5$), see Ref. \cite{Everts:2016b} for a detailed discussion. At smaller $d$, the image-charge interaction, which is attractive for both systems, becomes important. In the nanometer vicinity of the interface, the vdW repulsion dominates, and taken together with the image-charge potential, this gives rise to a minimum in {\mycolor{ $\Phi(d)\equiv H(d)-H(\infty)$}}, which corresponds to the equilibrium trapping distance of the particles from the interface.

Increasing $\kappa_o^{-1}$ reduces $|Z|$, such that the vdW repulsion can eventually overcome the image-charge potential for sufficiently small $d$ (Fig. \ref{fig:collint1}(a),(c)). However, the reduction in the particle-ion force is much smaller than the reduction of the image force, since the former scales like $\sim Z$, unlike the latter, which scales (approximately) like $\sim Z^2$. In Fig. \ref{fig:collint1}(a), we find that this results in a trapped state near the interface which becomes metastable for large $\kappa_o^{-1}$, with a reduced energy barrier upon increasing $\kappa_o^{-1}$. For system 2, we find that $\Phi(d)$ becomes repulsive for all $d$ for sufficiently large $\kappa_o^{-1}$, because the attractive image charge and the attractive colloid-ion force are reduced due to particle discharging. This calculation shows that  particle detachment from the interface is possible by removing a sufficient number of ions from the oil phase. This effect is stronger in system 1, because the repulsive Donnan-potential mechanism is longer ranged than the vdW repulsion. However, to the best of our knowledge, such detachment was \emph{not} observed in experiments by, for example, adding a sufficient amount of water that acts as an ion sink. Taken together with the experimental observation that initially positively charged particles can acquire a negative charge, we conclude that systems 1 and 2 are not described by single-adsorption models \cite{Elbers:2016}.

With the same procedure as for the single adsorption model, we determined the values of the equilibrium constants in the case of a  binary adsorption model. For system 1 we also used the salt concentration $\rho_{\text{TBA}^+}\big{|}_{Z=0}$ for which charge inversion takes place, to find $a^3K_{\mathrm{H}^+}=0.0001$, $a^3K_{\text{Br}^-}=47$, and $\theta=0.8$. Here $\theta=\theta_{\text{Br}^-}$ is the fraction of sites on which anions can adsorb. For system 2, we assumed $\theta=0.5$ and found $a^3K_{\mathrm{H}^+}=1$ and $a^3K_{\text{Br}^-}=0.055$. The short-distance (vdW), mid-distance (image charge) and long-distance (Donnan) behaviour of $\Phi(d)$ does not qualitatively change in the binary adsorption model, see Fig. \ref{fig:collint1}(b) and (d). However, the trapped state is more ``robust'' to changes in the ionic strength, because of the much higher values of $|Z(d)|$. This can be understood as follows. In system 1, $K_{\text{Br}^-}>K_{\text{H}^+}$, and thus decreasing the salt concentration leads the negatively charged surface sites to discharge first, which means that the charge initially increases with $\kappa_o^{-1}$. This enhances the image-charge effects, giving rise to a deeper potential well for the trapped state. At even higher $\kappa_o^{-1}$, $|Z(d)|$ will eventually decrease due to cationic desorption, although this is not explicitly shown in Fig. \ref{fig:collint1}. A similar reasoning applies to the negatively charged colloidal particles in system 2, which show only discharging upon increasing $\kappa_o^{-1}$, but much less compared to the single adsorption model.

The theoretically predicted stronger trapping in both systems and the experimentally observed sign reversal of the colloidal particles of system 1, which requires at least two adsorbed ionic species, indicates that the binary adsorption model describes the experiments better than the single adsorption model. In addition, the large energy barrier between the trapped state and the bulk in Fig. \ref{fig:collint1}(b), shows that not all the colloidal particles can be trapped near the oil-water interface. This is consistent with the experimentally observed zone void of colloidal particles, although one should keep in mind that the charged monolayer will provide additional repulsions which are not taken into account in our single-particle picture. 

\begin{figure*}
\includegraphics[width=0.8\textwidth]{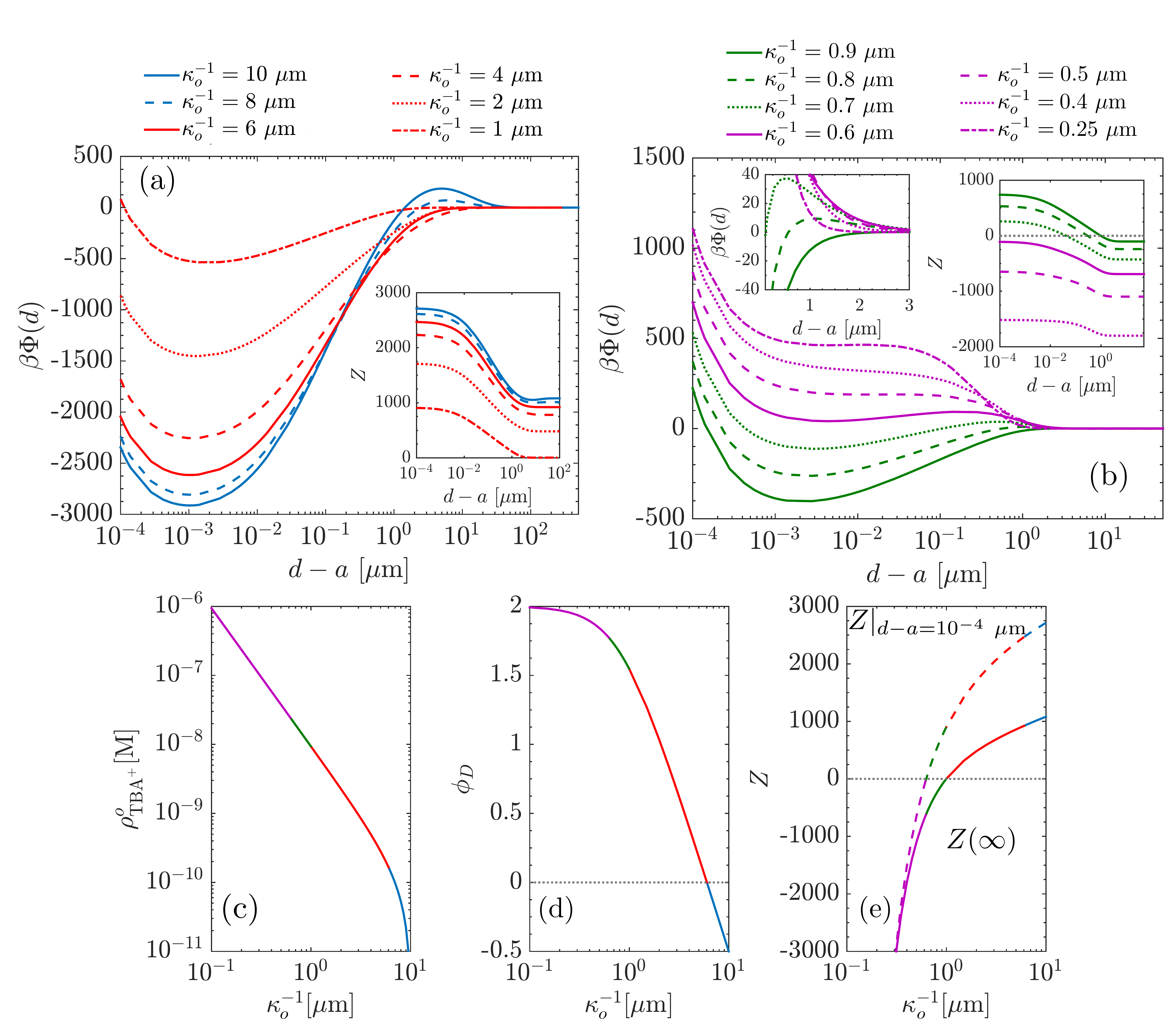}
\centering
\caption{Colloid-interface interactions for a three-ion model, in which only the ions $\mathrm{H}^+$ and $\mathrm{Br}^-$ can adsorb with the same equilibrium constants as in Fig. \ref{fig:collint1}(b). By fixing $\rho_{\text{H}^+}^o$ to the value of the blue full line where $\rho_{\text{TBA}^+}^o=0$ (which is the same as in Fig. \ref{fig:collint1}(b)), we show the effect of a decreasing screening length by the addition of $\rho_{\text{TBA}^+}^o$ in (a) and (b) with the particle charge $Z$ shown in the insets and a zoomed-in version of $\Phi(d)$ {\mycolor{(in units of $\beta^{-1}=k_BT$)}} in (b) to show more clearly the small maxima that are found for $\kappa_o^{-1}=0.7\ \mu\mathrm{m}$ and $0.8\ \mu\mathrm{m}$. In (c) we show how the resulting $\kappa_o^{-1}$ behaves as function of $\rho_{\text{TBA}^+}^o$, and changing this density does not only influence the Donnan potential $\phi_D/\beta e$ as shown in (d), but also (e) the charge $Z$ close to the interface as shown as the dashed line in and in the bulk oil as shown by the full line, because more $\text{Br}^-$ is available for adsorption. We use different colors to indicate the various regimes: blue is used for $\phi_D<0$ and $Z>0$, red for $\phi_D>0$ and $Z>0$, green for $\phi_D>0$ and $Z<0$ sufficiently far from the interface and purple for $\phi_D>0$ and $Z<0$ for all $d$.}
\label{fig:collint2}
\end{figure*}

\subsection{Systems with TBAB added}
We now show how the colloid-interface interaction changes in a system with three ionic species. We focus on the binary adsorption model applied to system 1, because this system has the richest behaviour, allowing $Z$ to switch sign. Here, the addition of TBAB gives rise to two new features. The first one is that it is possible to independently tune $\rho_{\text{Br}^-}^o$ and $\rho_{\text{H}^+}^o$ in the bulk oil phase while satisfying the constraint of bulk charge neutrality, $\rho_{\text{TBA}^+}^o+\rho_{\text{H}^+}^o=\rho_{\text{Br}^-}^o$. By increasing $\rho_{\text{TBA}^+}^o$, we find that $Z$ switches sign at
\begin{equation}
\rho_{\text{TBA}^+}\big{|}_{Z=0}=\frac{K_{\text{Br}^-}(1-\theta)\rho_{\text{H}^+}^o}{(2\theta-1)\rho_{\text{H}^+}^o+\theta K_{\text{H}^+}}, \label{eq:chargeswitch}
\end{equation}
where we used Eq. \eqref{eq:Langmuir} together with the condition $\sigma_{\text{Br}^-}=\sigma_{\text{H}^+}$. Secondly, because of the hierarchy $f_{\text{TBA}^+}<f_{\text{Br}^-}<f_{\text{H}^+}$, the Donnan potential can switch sign at
\begin{equation}
\rho_{\text{TBA}^+}\big{|}_{\phi_D=0}=\rho_{\text{H}^+}^o\frac{e^{f_{{\text{H}^+}}-f_{{\text{Br}^-}}}-1}{1-e^{f_{{\text{TBA}^+}}-f_{{\text{Br}^-}}}},\label{eq:donnanswitch}
\end{equation}
where we used Eq. \eqref{eq:donnan} and \eqref{eq:salt}. Eq. \eqref{eq:donnanswitch} is weakly dependent on the precise value of $f_{\text{TBA}^+}$, since $\exp(f_{\text{TBA}^+}-f_{\text{Br}^-})<0.02$ for $f_{\text{TBA}^+}\lesssim 6$  (with 6 being its value within the Born approximation), and hence the second term in the denominator of Eq. \eqref{eq:donnanswitch} can be neglected. Using the equilibrium constants of Sec. \ref{sec:eqnotbab}, we see from Eq. \eqref{eq:chargeswitch} and \eqref{eq:donnanswitch} that $\phi_D$ switches sign before $Z$ does upon adding TBAB; i.e., $\rho_{\text{TBA}^+}\big{|}_{\phi_D=0}<\rho_{\text{TBA}^+}\big{|}_{Z=0}$.

Since our calculations are performed in the grand-canonical ensemble, we have to specify how we account for the added TBAB. We choose to fix $\rho_{\text{H}^+}^o$, and set $\kappa_o^{-1}=10\ \mu\mathrm{m}$ without added TBAB (blue curve in Fig. \ref{fig:collint1}(b)). {\mycolor{The Debye length is chosen to be slightly larger than that of a pure CHB system, because the water phase acts as an ion-sink, see the discussion in Sec IV.}} The resulting colloid-interface interactions are shown in Fig. \ref{fig:collint2}(a) and (b), for various values of $\kappa_o^{-1}$, which decreases upon addition of TBAB. The relation between the screening lengths and the bulk concentration $\rho_{\text{TBA}^+}^o$ is shown in Fig. \ref{fig:collint2}(c). We can identify four regimes, indicated by different colors in Fig. \ref{fig:collint1}. We start with a system for which $\phi_D<0$ and $Z>0$ (blue curves), such that an energy barrier is present that separates the trapped state from the bulk state. Increasing $\rho_{\text{TBA}^+}^o$ decreases $|\phi_D|$ until ultimately the energy barrier vanishes and $\phi_D$ becomes positive (red curves). At even larger TBAB concentration, the colloidal particle becomes negative for $d\rightarrow\infty$ as it would be in bulk at the given $\kappa_o^{-1}$ (green curves). 

Interestingly, there is a (small) energy barrier of a different nature than the energy barriers shown until now. Namely, there exists a $d^*$ for which $Z(d^*)=0$ (see insets in Fig. \ref{fig:collint2}(b)). Surprisingly, at this point of zero charge, $d^*$ does not coincide with the location of the maximum in $\Phi(d)$. Furthermore, the result for $\kappa_o^{-1}=0.9$ $\mu$m does not show a maximum, although there is a point of zero charge. Both observations can be understood from the fact that although $Z=0$, the charge density $\sigma(\vartheta)$ is not spatially constant. In this case, there is still a coupling between bulk and surface ions, that contributes to $\Phi(d)$, see second term in Eq. \eqref{eq:effHam}.

Lastly, at a very high TBAB concentration we find $Z(d)<0$ for all $d$ (purple curves), and the large Donnan potential leads in a repulsion for all $d$, and hence to particle detachment. Upon decreasing $\kappa_o^{-1}$, this repulsion first becomes stronger, as $\phi_D$ increases towards $2$. At the same time, increasing $|Z|$  increases the strength of the image-charge attraction, eventually resulting in a plateau in $\Phi(d)$ between $d-a\sim10^{-3}\ \mu\mathrm{m}$ and $d-a\sim10^{-1}\ \mu\mathrm{m}$ (compare $\kappa_o^{-1}=0.25\ \mu\mathrm{m}$ with $\kappa_o^{-1}=0.4\ \mu\mathrm{m}$ in Fig. \ref{fig:collint2}(b)). 

We now briefly explain how added TBAB would change the colloid-interface interactions in the other cases presented in Fig. \ref{fig:collint1}(a), (c) and (d). In the case of a single adsorption model of system 1 only the Donnan potential switches sign, the energy barrier would vanish and the particles stay trapped. Possibly, some of the particles from the bulk are then moved towards the oil-water interface. For system 2, the addition of TBAB would only introduce an energy barrier separating the trapped state from a bulk state, but no detachment occurs, independent of the investigated adsorption model. This is in line with the experiments of Ref. \cite{Elbers:2016}, where no particle detachment was observed for system $2$. 

From the calculations in Fig. \ref{fig:collint2}, we deduce that significant particle detachment from the interface occurs whenever $Z<0$ and $\phi_D>0$. However, the range of the repulsion, which extends up to $1 \ \mu\mathrm{m}$, is too short to explain the particle detachment found in experiments, which may extend up to $>10 \ \mu\mathrm{m}$. One possible explanation to this discrepancy is that the particle motion far from the interface is governed by a non-equilibrium phenomenon, e.g. from the concentration gradient of ions generated by their migration from the oil phase to the water phase, similar to the recent experiment by Banerjee \textit{et al.} \cite{Squiresa:2016}. This motivated us to investigate the ion dynamics in the next section, in order to gain insight into the time evolution of the colloid-ion forces.

\section{Ion dynamics}
\label{sec:PNP}
For simplicity, we assume now that no colloidal particle is present in the system, such that the ion dynamics can be captured within a planar geometry. This can still give insight into the colloid-ion potential, because we deduced in our previous work that $\Phi(d)$ can be approximated by{\mycolor $\beta\Phi(d)\approx Z(\infty)\phi_0(d)$ for sufficiently large $d$}, with $\phi_0$ the dimensionless potential \emph{without} the colloidal particle \cite{Everts:2016b}. The theory can be set up from Eq. \eqref{eq:dft}, with the second line set equal to zero, and one should also keep in mind that $\mathcal{R}$ is the total system volume in this case. It is then possible to derive equations of motion for $\rho_{i,\pm}({\bf r},t)$ by using dynamical density functional theory (DDFT) \cite{Marconi:1999}. For ionic species $i$ with charge $\alpha=\pm$, the continuity equation reads
\begin{equation}
\frac{\partial\rho_{i,\alpha}({\bf r},t)}{\partial t}=-\nabla\cdot{\bf j}_{i,\alpha}({\bf r},t),
\label{eq:cont}
\end{equation}
with particle currents ${\bf j}_{i,\alpha}({\bf r},t)$ equal to 
\begin{align}
&{\bf j}_{i,\alpha}({\bf r},t)= \label{DDFT} \\
\!\!&-D_{i,\alpha}({\bf r})\rho_{i,\alpha}({\bf r},t)\nabla\left(\left. \frac{\delta(\beta\mathcal{F})}{\delta\rho_{i,\alpha}({\bf r})}\right\vert_{\substack{\rho_{i,\alpha}({\bf r},t)}}\!\!\!\!\!\!\!\!\!\!\!+\beta V_{i,\alpha}({\bf r})\right). \nonumber
\end{align}
Explicitly working out the functional derivative gives
\begin{align}
&{\bf j}_{i,\pm}({\bf r},t)=  \label{eq:current} \\
&\!\!\!-D_{i,\pm}({\bf r})\left\{\nabla\rho_{i,\pm}({\bf r},t)+\rho_{i,\pm}({\bf r},t)\nabla[\pm\phi({\bf r},t)+\beta V_{i,\pm}({\bf r})]\right\} \nonumber, 
\end{align}
with $D_{i,\alpha}(z)=(D_{i,\alpha}^o-D_{i,\alpha}^w)\Theta(z)+D_{i,\alpha}^w$, with $D_{i,\alpha}^o$ ($D_{i,\alpha}^w$) the diffusion coefficient of an ion of sign $\alpha$ in bulk oil (water).  Here, we have used the Einstein-Smoluchowski relation to relate the electric mobility to the diffusion constant. The time-dependent electrostatic potential $\phi({\bf r},t)$ satisfies the Poisson equation (neglecting retardation), 
\begin{align}
\nabla\cdot[\epsilon({\bf r})\nabla&\phi({\bf r},t)]/\epsilon_o=  \label{eq:poisson}  \\ 
&-4\pi\lambda_B^o\left[\sum_{j=1}^{N_+}\rho_{j,+}({\bf r},t)-\sum_{j=1}^{N_-}\rho_{j,-}({\bf r},t)\right]. \nonumber
\end{align}
Eqs. \eqref{eq:cont}-\eqref{eq:poisson} are the well-known the Poisson-Nernst-Planck equations, and we solve them under the boundary conditions
\begin{align}
\left.
\begin{array}{l l}
{\bf n}\cdot{\bf j}_{i,\alpha}({\bf r},t)=0 \\
{\bf n}\cdot\nabla\phi({\bf r},t)=0
\end{array}
\right\} \forall{\bf r}\in\partial \mathcal{R},\ \forall t\in[0,\infty), \label{eq:PNPbc}
\end{align}
which follow from global mass and charge conservation, respectively. 

We estimate the diffusion coefficients by making use of the Stokes-Einstein relation $D_{i,\pm}^j=(6\pi\beta\eta_j a_{i,\pm})^{-1}$, where $\eta_j$ the viscosity of the solvent ($j=o,w$). At room temperature we have $\eta_w=8.9\cdot 10^{-4}$ Pa$\cdot$s, while for CHB $\eta_o=2.269\cdot 10^{-3}$ Pa$\cdot$s. From these values we find:  $D_{\text{H}^+}^w=8.76\cdot 10^{-10}$ m$^2/$s, $D_{\text{TBA}^+}^w=4.54\cdot 10^{-10}$ m$^2/$s,       
 $D_{\text{Br}^-}^w=7.43\cdot 10^{-10}$ m$^2/$s, $D_{\text{H}^+}^o=3.44\cdot 10^{-10}$ m$^2/$s, $D_{\text{TBA}^+}^o=1.78\cdot 10^{-10}$ m$^2/$s and $D_{\text{Br}^-}^o=2.91\cdot 10^{-10}$ m$^2/$s.

\begin{figure*}
\includegraphics[width=0.9\textwidth]{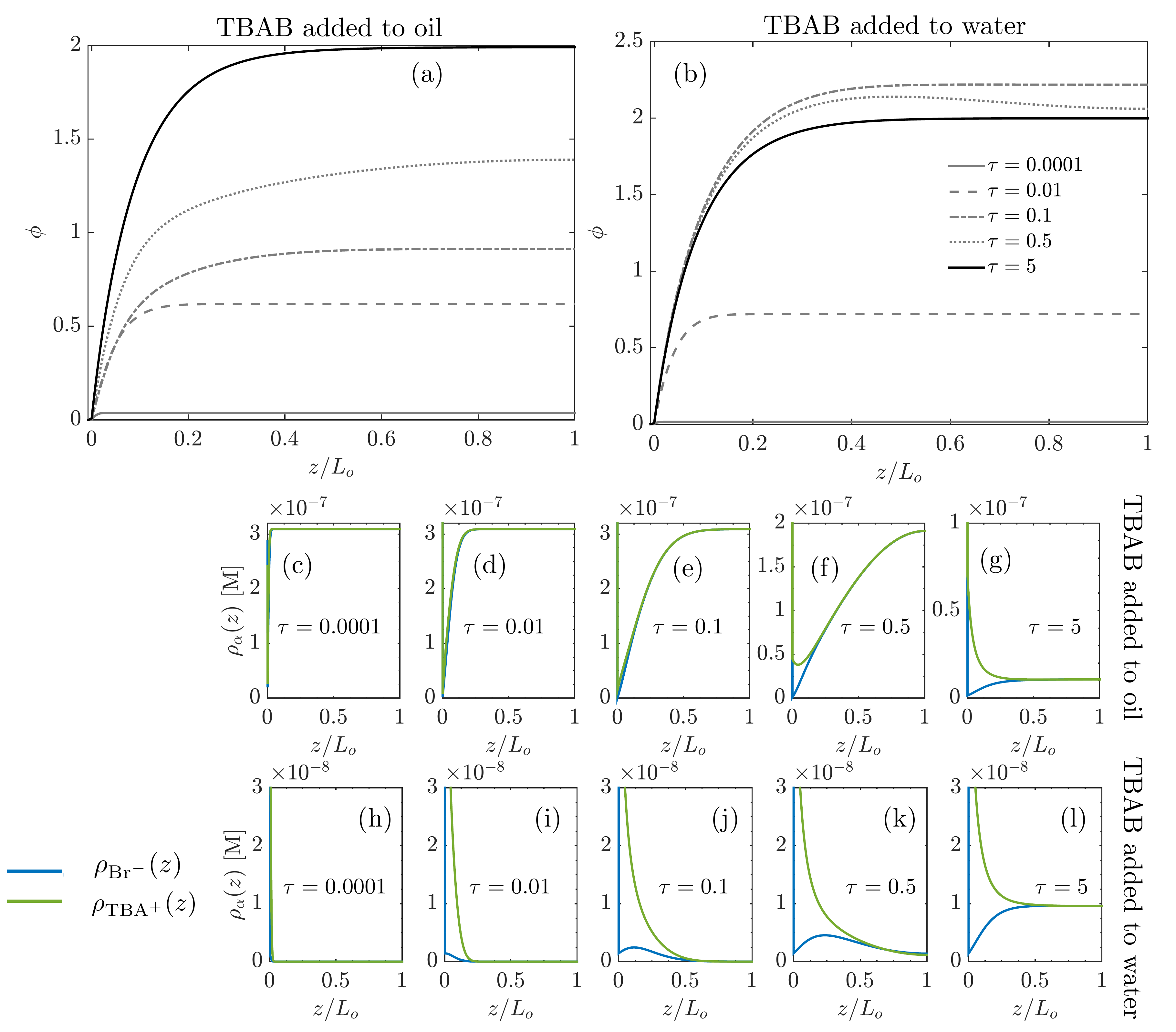}
\centering
\caption{Time evolution of a system with the ions TBA$^+$ and Br$^-$ in a capillary filled with a layer of oil with thickness $L_o=10\ \mu$m and a layer of water with $L_w=0.1\ \mu\mathrm{m}$. We show (a) the electrostatic potential $\phi(z,t)/(\beta e)$ in case the TBAB salt is added (a) to the oil with initially no ions in water, and (b) to water with initially no ions in the oil. The concentration profiles $\rho_{\text{TBA}^+}(z,t)$ and $\rho_{\text{Br}^-}(z,t)$, when TBAB is added to oil ((c)-(g)) and to water ((h)-(l)), are all shown in terms of the dimensionless time $\tau=t/t_0$, with $t_0=L_o^2/D_{\text{Br}^-}=1.3$ s. For $t\rightarrow\infty$ the screening length in oil is $\kappa_o^{-1}=0.979\ \mu\mathrm{m}$ in both cases. We do not show the concentration profiles in water because they were unrealistic in the model that we used (see main text). Finally, note that in all of our calculations $\tau=5$ is essentially the same as $\tau\rightarrow\infty$.}
\label{fig:TBABoil}
\end{figure*}

\subsection{Dynamics after TBAB addition}

The ion dynamics can provide further insight into the particle dislodgement after TBAB is added to the oil phase. 
In experiment, we observed that $\kappa_o^{-1}$ can be decreased down to 50 nm, after TBAB is added. This Debye length implies a salt concentration of the order of $ 10^{-7}$ M, such that we can safely neglect the HBr concentration, which has a maximal value of $\sim 10^{-10}$ M before the oil is brought into contact with the water phase. 

We investigate the time-dependence of the electrostatic potential $\phi(z,t)$, with $z$ the direction perpendicular to the oil-water interface. The oil is assumed to reside in a capillary with a linear dimension perpendicular to the oil-water interface of length $L_o=10\ \mu\mathrm{m}$, which is much larger than $\kappa_o^{-1}$ but much smaller than the experimental sample size of about $1$ cm, to facilitate numerical calculations. It was difficult to perform calculations at even larger $L_o$ with such a small $\kappa_o^{-1}$, but the present parameter settings can nevertheless give qualitative insights. In experiments, the length of the water side of the capillary $L_w$ is also 1 cm, but here we take it to be $L_w=0.1\ \mu\mathrm{m}$, which is still much larger than $\kappa_w^{-1}$. The disadvantage of the small $L_w$ is that only the ionic profiles in the oil phase are considered realistic, because given the small $L_w$ no bulk charge neutrality in the water phase can be obtained. {\mycolor Furthermore, $L_w\ll L_o$ stems from the initial condition that we define below together with the desired final condition, constrained by the fact that ions cannot leave the oil-water system and that the water phase is modeled as an ion-less ion sink. In contrast, for the calculation of the effective colloid-oil-water-interface potential in Sec. IV, we used a grand-canonical treatment, rather than a canonical treatment for the ions that is used for the dynamics here.}


Similar to the experiments, the initial condition for $(i,\alpha)=\mathrm{TBA}^+, \mathrm{Br}^-$ is a uniform distribution of ions in the oil phase:
\begin{equation}
\rho_{i,\alpha}(z,t=0)=\rho_0\Theta(z). \label{eq:PNPin}
\end{equation}
The amplitude $\rho_0=[\kappa_o(t=0)]^2/8\pi\lambda_B^o$ is used such that we can acces the regime where the particles are negatively charged for $d\rightarrow\infty$ and $t\rightarrow\infty$, but they can become positively charged close to the interface. In particular, we use $\kappa_o^{-1}(t=0)=0.05\ \mu\mathrm{m}$, leading to a final $\kappa_o^{-1}(t\rightarrow\infty)=0.979\ \mu\mathrm{m}$ (cf. Fig. \ref{fig:collint2}(b)). Solving Eq. \eqref{eq:cont}, \eqref{eq:current}, \eqref{eq:poisson}, with boundary conditions \eqref{eq:PNPbc} and initial condition \eqref{eq:PNPin}, results in the profiles $\phi(z,t)$, $\rho_{\text{H}^+}(z,t)$, and $\rho_{\text{Br}^-}(z,t)$. It is convenient to express the results in terms of a dimensionless time $\tau=t/t_0$, with time scale $t_0=L_o^2/D_{\text{Br}^{-}}^w$, which in our system is $t_0=1.3$ s. This means that the equilibrium state is reached within several seconds in our system, see the profiles in Fig. \ref{fig:TBABoil}. However, if a more realistic $L_o$ is chosen, this time scale will be on the order of hours, since $t_0$ scales with $L_o^2$.

In Fig. \ref{fig:TBABoil}(a), we show the time evolution towards equilibrium of $\phi(z,t)$. For all times, $\phi(z,t)$ increases monotonically with $z$ and becomes constant as $z\rightarrow L_o$. The range of $\phi(z,t)$ steadily increases over time due to the depletion of ions in the oil. In addition, $\phi(L_o,t)$ increases with time, until ultimately $\phi(L_o,t\rightarrow\infty)=\phi_D$ is reached.


The equilibrium calculations of Fig. \ref{fig:collint2} supported particle detachment by means of a repulsive colloid-ion force, but due to the large salt concentrations the range of the repulsive colloid-ion force was deemed to be too small in the parameter regime where the particle was negatively charged. The dynamics of the ionic profiles at the oil side, presented in Fig. \ref{fig:TBABoil}(c)-(g)), show that this issue can be resolved when the system is (correctly) viewed out of equilibrium, as we will explain next.

From the profiles in Fig. \ref{fig:TBABoil}(c), a short time after the addition of salt, we infer that the colloids are initially negatively charged according to the corresponding $\kappa_o^{-1}$ and $Z$ in Fig. \ref{fig:collint2}(c). Therefore, the approximate interaction potential $\beta\Phi(d)\approx Z(\infty)\phi_0(d)$ leads to a colloid-ion force that is repulsive. Colloidal particles that were initially trapped are then repelled from the interface, but only for surface-interface distances up to a micron, as can be inferred from Fig. \ref{fig:collint2}(b). When $t$ increases, the water phase uptake of ions reduces the Br$^-$ concentration close to the interface. At the same time, mass action is at play, and we can estimate from \ref{fig:collint2}(c) that the particles become positively charged at $\approx 10^{-8}$ M. This means that as time progresses, some of the particles close to the interface will reverse their sign. For example, at time $\tau=0.5$, we can estimate from the profiles in \ref{fig:TBABoil}(f) that only particles at $d\gtrsim1 \ \mu\mathrm{m}$ are still negatively charged. However, assuming that the bulk ion dynamics is much slower than the mass action dynamics, the range of the Donnan potential has not relaxed yet, and is longer ranged than at $t\rightarrow\infty$. At $\tau=0.5$, $\phi$ still extends up until $L_o=10\ \mu\mathrm{m}$, see the dotted line in Fig. \ref{fig:TBABoil}(a). Hence, the range of repulsion for the negatively charged particles is longer than one would expect from the equilibrium calculation. 
In other words, the range of the interaction is set much faster than the electrostatic potential and the colloidal charge at large $z$. At later times, enough ions are depleted from the oil, all the colloids become positively charged, and are attracted towards the interface, as one would expect in equilibrium for the final $\kappa_o^{-1}$. This also gives a possible explanation for the experimentally observed reattachment after the initial detachment.

For comparison, we also performed calculations with HBr as the only salt (no added TBAB). We found that except at the very early stages of the dynamics, the HBr concentration is indeed negligible and decreases rapidly after the oil comes into contact with the water due to the ion partitioning. These calculations also confirmed that, within the binary adsorption model, the colloid-ion forces remain repulsive throughout the partitioning processes, since particles becomes more positively charged with decreasing the ionic strength, because of the larger desorption of negative ions than positive ions. Thus, the colloid-interface interaction is still always dominated by the attractive short range image forces.

Finally, we consider what happens when TBAB is added to the water, neglecting the HBr concentration. In \ref{fig:TBABoil}(b), we show $\phi(z,t)$, and find that the potential in this case can temporarily become larger than $\phi_D$. The ion densities behave as expected. Some of the ions from the water side are transferred towards the oil phase. In \ref{fig:TBABoil}(h)-(l) we see that the density of ions is first largest at the interface until, slowly, also the rest of the oil is filled. Note that the oil side of the interface is always positively charged, and that the equilibrium situation is identical to the one in Fig. \ref{fig:TBABoil} by construction. Based on the calculation of Fig. \ref{fig:TBABoil}(b), we conclude that the colloid-ion forces are attractive for all times up until equilibrium is nearly reached. Because there is a high density of Br$^-$ ions in bulk, the particles are negatively charged sufficiently far from the interface. The colloids for small $d$ are, however, positively charged as was explained in the inset of Fig. \ref{fig:collint2}(b) (green curves). This explains why colloids are drawn closer to the interface upon adding TBAB in water: the colloids remain mainly positive, but a positive Donnan potential is generated out of a negative one, and hence an attraction towards the interface is induced. This we have already understood from the equilibrium calculations. 

\subsection{Diffusiophoresis}
Despite having only discussed electrostatic forces generated by the Donnan potential, our calculations can also give some insight into diffusiophoretic effects, that is, those induced by the motion of colloidal particles in concentration gradients of ions. We now estimate the importance of diffusiophoresis in both the HBr and added TBAB systems using the PNP calculations. Whenever the unperturbed concentration fields satisfy $\rho_+(z)\approx\rho_-(z)$, a negligble electric field is generated by the ions that would give rise to the aforementioned colloid-ion force. However, in an overall concentration gradient, the particles can be translated due to diffusiophoresis, in which the particle velocity is given by ${\bf U}=b \nabla[\rho_+(z)+\rho_-(z)]$, with slip-velocity coefficient
\begin{equation}
b=\frac{4k_BT}{\eta_o\kappa_o^2}\left\{\frac{\zeta}{2}\frac{D_+-D_-}{D_++D_-}-\ln\left[1-\tanh^2\left(\frac{\zeta}{4}\right)\right]\right\},\label{eq:slip}
\end{equation}
see Ref. \cite{Anderson:1989} for details. Note that Eq. \eqref{eq:slip} is derived assuming a homogeneous surface potential $\phi_0$, and that only the gauged potential $\zeta=\phi_0-\phi_D$ is relevant for an oil-dispersed colloidal particle.

From Eq. \eqref{eq:slip}, we can estimate the sign of $b$. For a system that contains HBr only, we find $b>0$, and hence colloidal particles tend to always move towards higher concentrations. This means that diffusiophoresis repels particles from the interface, similar to the colloid-ion force that we described in equilibrium. We therefore conclude that without TBAB, attractions are provided solely by the image charge forces.

When TBAB is added, we find that $b\leq0$ for $0\leq\zeta\lesssim2$ and $b>0$ otherwise. For TBAB in oil, the negatively charged particles therefore experience a repulsive diffusiophoretic force from the oil-water interface, while positively charged particles are attracted for $0\leq\zeta\lesssim 2$, but are repelled otherwise. Assuming that for TBAB in water the particles are always positively charged, particles with $\zeta>2$ are attracted to the interface by diffusiophoresis. Given that the particles in our studies were (relatively) highly charged, all forces except for the vdW (image charge, colloid-ion and diffusiophoretic force) are attractive in this specific case.

We conclude that diffusiophoresis could possibly account for the long range repulsion or attraction near the oil-water interface, since concentration gradients occur over a scale that is much larger than the Debye screening length. In fact it could suggest that diffusiophoresis is the dominant force generating mechanism outside of the double layer near the oil-water interface. However, the equilibrium considerations in Sec. \ref{sec:collint} are pivotal to understanding why colloidal particles can be detached in the first place.

\section{Conclusion and outlook}
In this paper, we discussed colloid--oil-water-interface interactions and ion dynamics of PMMA colloids dispersed in a non-polar oil at an oil-water interface, in a system with up to three ionic species. We have applied a formalism that includes ion partitioning, charge regulation, and multiple ionic species to recent experiments \cite{Elbers:2016}, to discuss (i) how the charges on the water and oil side of the oil-water interface can change upon addition of salt, (ii) how charge inversion of interfacially trapped non-touching colloidal particles upon addition of salt to the oil phase can drive particles towards the bulk over long distances, followed by reattachment for large times, (iii) that particles that cannot invert their charge stay trapped at the interface, and (iv) that colloids in bulk can be driven closer to the interface by adding salt to the water phase. We used equilibrium and dynamical calculations to show that these phenomena stem from a subtle interplay between long-distance colloid-ion forces, mid-distance image forces, short-distance vdW forces, and possibly out-of-equilibrium diffusiophoretic forces. The colloid-ion forces are the most easily tunable of the three equilibrium forces, because they can be tuned from repulsive to attractive over a large range of interaction strengths. We have shown this explicitly by including three ionic species in the theory, and by investigating various charge regulation mechanisms, extending the formalism of Ref. \cite{Everts:2016b}. 

For future directions, we believe that it would be useful to investigate many-body effects, in a similar fashion as in Ref. \cite{Zwanikken:2007}. There are, however, two drawbacks of the method of Ref. \cite{Zwanikken:2007} that need to be amended before we could apply it to a system of non-touching colloids. First of all, in Ref. \cite{Zwanikken:2007}, a Pieranski potential \cite{Pieranski:1980} was used to ensure the formation of a dense monolayer at the oil-water interface. It would be interesting to see if the trapping of particles near the interface can be found self-consistently by the mechanism presented here and the one of Ref. \cite{Zwanikken:2007}, by using a repulsive vdW colloid-interface potential. Secondly, the formalism of Ref. \cite{Zwanikken:2007} was set up for constant-charge particles. In the constant-charge case, it is a good approximation to replace the particle nature of the colloids by a density field. For charge-regulating particles, this can be a limiting approximation because one needs the surface potential and not the laterally averaged electrostatic potential to determine the colloidal charge. 

Investigating many-body effects can be interesting, because colloidal particles present in bulk contribute to the Donnan potential. This is not the case when all the colloids are trapped near the interface: in this case the electrostatic potential generated by the colloids cannot extend through the whole system volume. Finally, a dense monolayer can provide an additional electrostatic repulsion for colloids, in addition to the repulsive colloid-ion force for $Z(\infty)\phi_D<0$ and the repulsive vdW force. Therefore, we expect that the interplay of the colloidal particles with ions can be very interesting on the many-body level, especially when we include not only image-charge and ion-partitioning effects, but most importantly, also charge regulation. However, it is not trivial to take all these effects into account in a many-body theory. Another direction that we propose is to perform the ion dynamics calculation of Sec. \ref{sec:PNP} in the presence of a single (and maybe stationary) charged sphere near an oil-water interface. This would give insights into the out-of-equilibrium charging of charge-regulating particles, providing more information on the tunability of colloidal particles trapped near a ``salty'' dielectric interface. 

We acknowledge financial support of a Netherlands Organisation for Scientific Research (NWO) VICI grant funded by the Dutch Ministry of Education, Culture and Science (OCW) and from the European Union's Horizon 2020 programme under the Marie Sk\l{}odowska-Curie grant agreement No. 656327. This work is part of the D-ITP consortium, a program of the Netherlands Organisation for Scientific Research (NWO) funded by the Dutch Ministry of Education, Culture and Science (OCW). J.C.E. performed the theoretical modelling and numerical calculations under supervision of S.S. and R.v.R. The experiments were performed by N.A.E. and J.E.S.v.d.H. under the supervision of A.v.B. The paper is co-written by J.C.E. and S.S., with contributions of N.A.E., J.E.S.v.d.H., A.v.B. and R.v.R. The supplemental information is provided by N.A.E. and J.E.S.v.d.H. All authors discussed results and revised the paper.

\bibliographystyle{apsrev4-1} 
\bibliography{literature1_JvdH_NE-2} 

\end{document}